\documentclass[conference]{IEEEtran}
\makeatletter
\def\ps@headings{%
\def\@oddhead{\mbox{}\scriptsize\rightmark \hfil \thepage}%
\def\@evenhead{\scriptsize\thepage \hfil \leftmark\mbox{}}%
\def\@oddfoot{}%
\def\@evenfoot{}}
\makeatother
\IEEEoverridecommandlockouts
\usepackage{cite}
\usepackage{graphicx}
\usepackage[cmex10]{amsmath}
\usepackage{amssymb}
\usepackage{url}
\usepackage{subfigure}
\usepackage{algorithmic}
\usepackage{algorithm}
\usepackage{array}
\usepackage{comment}
\usepackage{multirow}
\usepackage{flushend}
\usepackage{subfigure,textcomp}

\begin{document}

\title{Evaluating Opportunistic Delivery of Large Content with TCP over WiFi in I2V Communication}
\author{\IEEEauthorblockN{Shreyasee Mukherjee, Kai Su, Narayan	B. Mandayam, K. K. Ramakrishnan, Dipankar	Raychaudhuri and  Ivan Seskar}
\IEEEauthorblockA{WINLAB, Rutgers University\\
671 Route 1 South, North Brunswick, New Jersey 08902\\
Email: \em\{shreya, kais, narayan, kkrama, ray, seskar\}@winlab.rutgers.edu}
\vspace{-7mm}
\thanks{*K. Su and N. B. Mandayam are supported in part by the NSF under grant no. CCF-1016551. This work was presented in IEEE LANMAN workshop 2014.}
}

%
\maketitle
\vspace{-3cm}
\vspace{-6mm}
\begin{abstract}

With the increasing interest in connected vehicles, it is useful to evaluate
the capability of delivering large content over a WiFi infrastructure to vehicles. The throughput achieved over WiFi channels can be highly variable and also rapidly degrades as the distance from the access point increases. While this behavior is well understood at the data link layer, the interactions across the various protocol layers (data link and up through the 
transport layer) and the effect of mobility may reduce the amount of content transferred to the vehicle, as it travels along the roadway.

This paper examines the throughput achieved at the TCP layer over a carefully designed outdoor WiFi environment and the interactions across the layers that impact the performance achieved, as a function of the receiver mobility. 
The experimental studies conducted reveal that impairments over the WiFi link (frame loss, ARQ and increased delay) and the residual loss seen by TCP causes a cascade of
duplicate ACKs to be generated. This triggers large congestion window reductions at the sender,
leading to a drastic degradation of throughput to the vehicular client.
To ensure outdoor WiFi infrastructures have the potential to sustain reasonable downlink throughput for drive-by vehicles, 
we speculate that there is a need to adapt how WiFi and TCP (as well as mobility protocols) function for such vehicular applications.

\vspace{-2mm}
\end{abstract}


\section{Introduction}

There has been a growing interest in having connected vehicles, with the ability to communicate wirelessly at all times.
While much of the focus has been on Intelligent Transport Systems (ITS) to develop services for traffic alerts and safety applications \cite{safety-url, 2001-its},  
there is also a significant interest in delivering large entertainment oriented content and other
information to the vehicle. Much of this communication is likely to occur with the vehicle communicating with infrastructure
nodes, in what is known as Vehicle-to-Infrastructure (V2I) Communications.
While the adoption of cellular communications in the connected vehicle is beginning, it has been slow to grow because of the
current, high cost of cellular data communication to the consumer. On the other hand, delivering the
desired information over a WiFi infrastructure that may  already exist or can be deployed at a minimal cost on roadways, may be an attractive alternative.
However, it is important to achieve sufficiently high throughput for the data transfer because a moving vehicle is likely to be associated with any given access point (AP) only for short periods of time.

There have been a large number of efforts to provide public WiFi infrastructure in urban areas in the United States in recent years.
The primary focus has been to provide ubiquitous Internet access through these WiFi APs. When deploying  Municipal WiFi networks \cite{muni-map}, for example, ``Google WiFi" in Mountain View, California \cite{google-wifi}, 
access points are mounted on roadside
lampposts. Such outdoor WiFi AP deployments are designed primarily for pedestrian Internet access. The V2I infrastructures
may enhance these capabilities by having WiFi APs operate at points more suited to vehicular traffic, such as being deployed at
intersections, possibly at higher elevation (on traffic lights) etc. 
So, if there is a significant amount of WiFi deployed via a combination of Municipal WiFi, V2I, and other `freely' accessible WiFi hotspots, the question then to be addressed is: how useful will all these WiFi deployments be also for
delivering information to moving vehicles. In \cite{des-imc-10}, it was shown through repeated experiments of measuring WiFi throughput over a $9$-mile drive that 60\% of the time when speed was within $20$km/h, vehicles could intermittently achieve approximately 2Mbps throughput when downloading content over WiFi networks.

It is well-known that the throughput achieved over WiFi channels can be highly variable and also rapidly degrades as the distance from the AP increases. While this behavior is well understood at the data link layer, the interactions across the various protocol layers (data link, network layer as a result of mobility and transport layer) may eventually result in a substantially lower amount of aggregate content transferred to the vehicle as it travels along the roadway.


In this paper, we evaluate the effectiveness of I2V content delivery. 
We present measurement results from a set of carefully designed outdoor WiFi experiments  
emulating a variety of vehicular movement scenarios. For example when a client moves at various speeds (including a pedestrian walking speed), 
is stationary or stopped at different locations with respect to the AP. Our goal with these experiments is to understand
and identify the unique challenges \textit{mobility} poses to I2V communications over WiFi. This takes on particular
significance when the actions taken by the 802.11 MAC protocol and TCP interact in unpredictable ways. 
The concern with content delivery over a V2I infrastructure is that a vehicle is likely to be associated with a given AP for a brief 
period of time. Maximizing the amount of data delivered to the client during this short interval is key to the success of content delivery
in the vehicular context.
From our experimental results, we observe that a packet loss that is not related to congestion has the consequence of
causing TCP to prematurely drop the congestion window size substantially and correspondingly
degrading the throughput achieved. While there are several reasons for TCP to drop its window size to the initial value, including
timeouts, one of the characteristics we observed was the generation of a large number of DupAcks by the receiver after a packet was 
irrecoverably lost by the data link layer (e.g., after the number of link layer ARQs exceed the limit set in the 802.11 implementation).
The sender, after seeing a large number of DupAcks, drops the window to the minimum value of 1, and the throughput in turn degrades substantially 
even though the channel is uncongested. While this effect is tolerable in a continuous, and long-lived static connection, it 
is mandatory that this be avoided in I2V content delivery, as each vehicle's connection time is limited, often of the order of a few 
tens of seconds.  


In what follows, we first briefly introduce previous work on vehicular WiFi access. We then describe the physical setup and methodology of our experimental study of I2V data delivery performance over WiFi. We present our results in section IV, focusing on dissecting 802.11 MAC and TCP's respective mechanisms especially with regard to their reaction to loss, and analyze the impact on overall throughput for content delivery.
After that, we  discuss the implications of our experimental observations, and the possible network design choices to improve vehicular WiFi content delivery. 
\vspace{-2mm}


\section{Related Work}
There exist a body of previous work on experimental study of vehicular WiFi access performance, from 
a variety of perspectives.

Most prior work, such as~\cite{ott2004drive, 05gass} focus primarily on the feasibility and performance characteristics of V2I communication at different speeds and environments. In \cite{had-mobisys-07}, Hadler et al. provide experimental study results of a static AP transmitting to a drive-by vehicle moving at a speed of $80$km/h. They show that the current protocol stack achieves half of the available bandwidth, and focus on analyzing the overhead which causes throughput underutilization during connection establishment and data delivery. The topology adopted in our experiments involves the data sender as a third node, such that the interplay of TCP's end-to-end performance and the AP's 802.11 MAC layer reliability mechanisms can be studied. In a vehicular environment, clients possibly move at different speeds, and frequently lose association with a currently connected AP and need to re-associated with a new one. The study in \cite{bal-ccr-08} showed that by allowing the mobile client to opportunistically associate with multiple APs, one could avoid the overhead related to handoffs. This reduces connection disruptions and improves user experience. This approach would be another way of prolonging the vehicular WiFi connection time, which is beneficial for I2V communications. 
Recent works~\cite{hare2013dept} and ~\cite{hare2012} have also studied  using multiple 3G interfaces to improve data delivery to vehicular clients by splicing data over multiple 3G interfaces. While this is useful for V2I communication,
it is somewhat orthogonal to our focus of delivering content over WiFi. We
seek to more carefully understand the effectiveness of delivering content
over a single TCP connection over the WiFi interface. 

There has also been a significant amount of work in the ITS community to use Dedicated Short Range Communication (DSRC) for delivering small amounts of data with minimum delay~\cite{06-dsrc}. However, our focus in this study is to analyze the ability of the network to sustain high throughput for large content delivery services. A number of papers also compare and contrast WiFi and cellular access from vehicles, and consider offloading cellular data to WiFi so as to reduce usage cost. In \cite{bal-mobisys-10}, an experimental study of network performance, in terms of TCP throughput and loss rate, was carried out for both WiFi and 3G. The authors concluded that the median downlink WiFi throughput is less than half of 3G's, and WiFi has significantly higher loss rate than 3G. This paper suggests that WiFi's high data rate could be utilized to deliver delay-tolerant flows, thus reducing cellular data usage. 
In this work, we seek to first understand what are the causes for vehicular TCP over WiFi throughput degradation, so that future designs can take them into account.

\section{Experimental Methodology} 
\begin{figure}[!t]
\centering
\includegraphics[width=0.8\columnwidth]{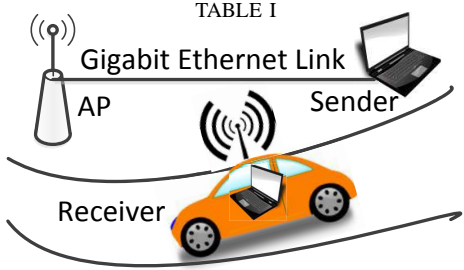}
\vspace{-0.5cm}
\caption{Experimental Topology}
\label{fig:topo}
\vspace{-0.5cm}
\end{figure}

%
We consider a simple setup of an outdoor, open, 802.11g WiFi Access Point delivering content to a vehicular client in a variety of conditions. We assume that content is delivered using TCP as the transport protocol. We examine the performance in such situations by transferring large
amounts of data between a node (`sender') connected to the AP and the mobile client (`receiver'), as shown in Fig.~\ref{fig:topo}. 
We evaluate the performance in this environment by using Iperf ~\cite{iperf} to transfer large (1300) byte packets from the sender to the receiver. The sender is an `Iperf client', connected to the AP with Gigabit Ethernet (so as to ensure that link is not a bottleneck) and the receiver is the `Iperf server' running on the mobile client. Both the sender and receiver nodes are specially built high performance nodes with an Intel dual core 2.8Ghz processor configured with ample memory so as to be able to sustain full link rate transmission/reception and
have enough processing capability to support wireshark/tcpdump/tcpprobe at line rate. The AP is also a node with the same processor. The WiFi AP and link are configured using hostapd to run 802.11g with a nominal link rate of 54 Mbps. We ran experiments by having the IEEE 802.11 auto-rate rate adaptation algorithm turned off.  
The AP was set to operate on channel 11, after ensuring that there were no other APs in the vicinity operating on that same channel. 
Since the AP is purposely built for experimental use, it was possible to increase the output power to examine the effectiveness of the outdoor AP to deliver content to a vehicle at larger
distances (greater than 100 meters). We began our experiments with the AP configured to deliver 500 milliwatts of transmit power and report a first set of measurements. 

After positioning the AP on the side of the road, we conducted several tests, the results of which we report in  this paper. 
\begin{itemize}
\item With the receiver moving at the speed of a typical pedestrian
\item The receiver moving slowly towards and past the AP, and going around a building that blocks the line of sight to the AP
\item The receiver moving from position to position, with a long stop at each position (to measure Iperf throughput)

\end{itemize}


The specifications of the nodes are summarized in table~\ref{table:specs}.

\begin{table}[t]
\vspace{-0.5cm}
\begin{tabular}{|c|c|}
\hline
\multirow{3}{*}{AP} & CPU: Intel(R) Core(TM)2, 2.80 GHz\\
\cline{2-2}
& Antenna: Omni-directional 6 dbi dipole\\
\cline{2-2}
& Wifi Driver: AR922X Wireless Network Adapter\\
\hline
\multirow{1}{*}{Sender} & CPU: Intel(R) Core(TM)2, 2.80 GHz \\
\hline\multirow{3}{*}{Receiver} & CPU: Intel(R) Core(TM)2, 2.80 GHz\\
\cline{2-2}
& Antenna: Omni-directional 6 dbi dipole\\
\cline{2-2}
& Wifi Driver: AR922X Wireless Network Adapter\\
\hline
\end{tabular}
\caption{}
\label{table:specs}
\vspace{-1cm}
\end{table}

The sender and receiver were running TCP-Reno (changed from the default setting of TCP-Cubic on Linux). Iperf was set to 
transmit continuously throughout each of the experiments, sending TCP packets with 1300 bytes as payload. TCP-SACK was enabled and the receive socket buffer was set to 64 Kbytes. The AP kernel buffer was set by default to have a 1000 frame buffer. 

The sender and receiver TCP behavior was monitored using a variety of tools that all relate to tcpdump.
TCP Probe, a Linux kernel module was run on the sending node to record the state of the TCP connection. In addition, the driver at another Ubuntu machine, used as a `sniffer' was configured to create a virtual interface that logged tcpdump at the AP throughout each experiment. Tcpdump was also run on the Ethernet port at the sender. We post-processed these packet dumps with tshark to obtain 802.11 MAC header, TCP headers, received signal strength, physical layer data rate along with timestamps for the experiments. We also logged the receiver position using GPS so as to relate the physical location as well as the speed of the receiver as it was driven in the various ways described above.
\vspace{-2mm}

\section{Experimental Results}
\begin{figure}[!t]
\vspace{-0.8cm}
\centering
\includegraphics[width=\columnwidth]{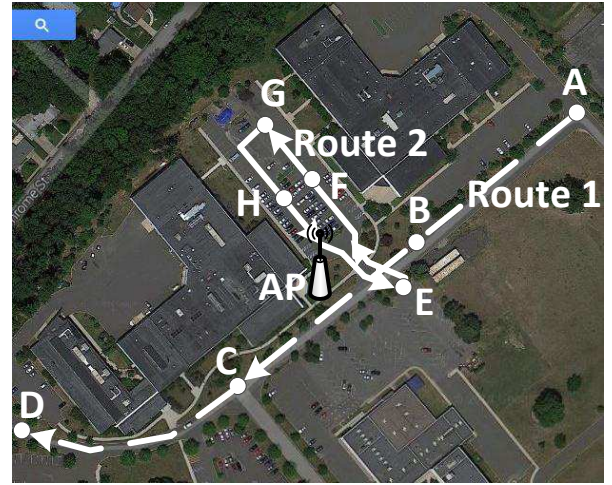}
\vspace{-0.8cm}
\caption{Physical routes followed by the receiver during the experiments}
\label{fig:exp_drive}
\vspace{-0.5cm}
\end{figure}

We present experimental results, focusing primarily on the three scenarios described earlier: (i) pedestrian walking from one end of a road to the other end, until the receiver loses line of sight (shown as ``Route 1" in Fig. \ref{fig:exp_drive}); (ii) receiver driving slowly around a parking lot; it loses
line of sight with the AP for an interval (shown as ``Route 2" in Fig. \ref{fig:exp_drive}); (iii) 
receiver remains stationary (at the point ``B" in Fig.~\ref{fig:exp_drive}). 
For the first two experiments, at the starting point, the receiver performs association and authentication with AP, and obtains an IP address assigned by DHCP. Only then the receiver starts an Iperf server to listen on the default port of $5001$. After that, the receiver starts moving. For all three
experiments we record the 
statistics for the experiment after all the initial TCP connection is setup, to correctly estimate the
TCP throughput.  

\vspace{-4mm}
\subsection{Pedestrian scenario}
\vspace{-1mm}

\begin{figure*}[!t]
\vspace{-0.5mm}
\hspace{-1cm}
\subfigure[]{
\includegraphics[width= 0.33\linewidth]{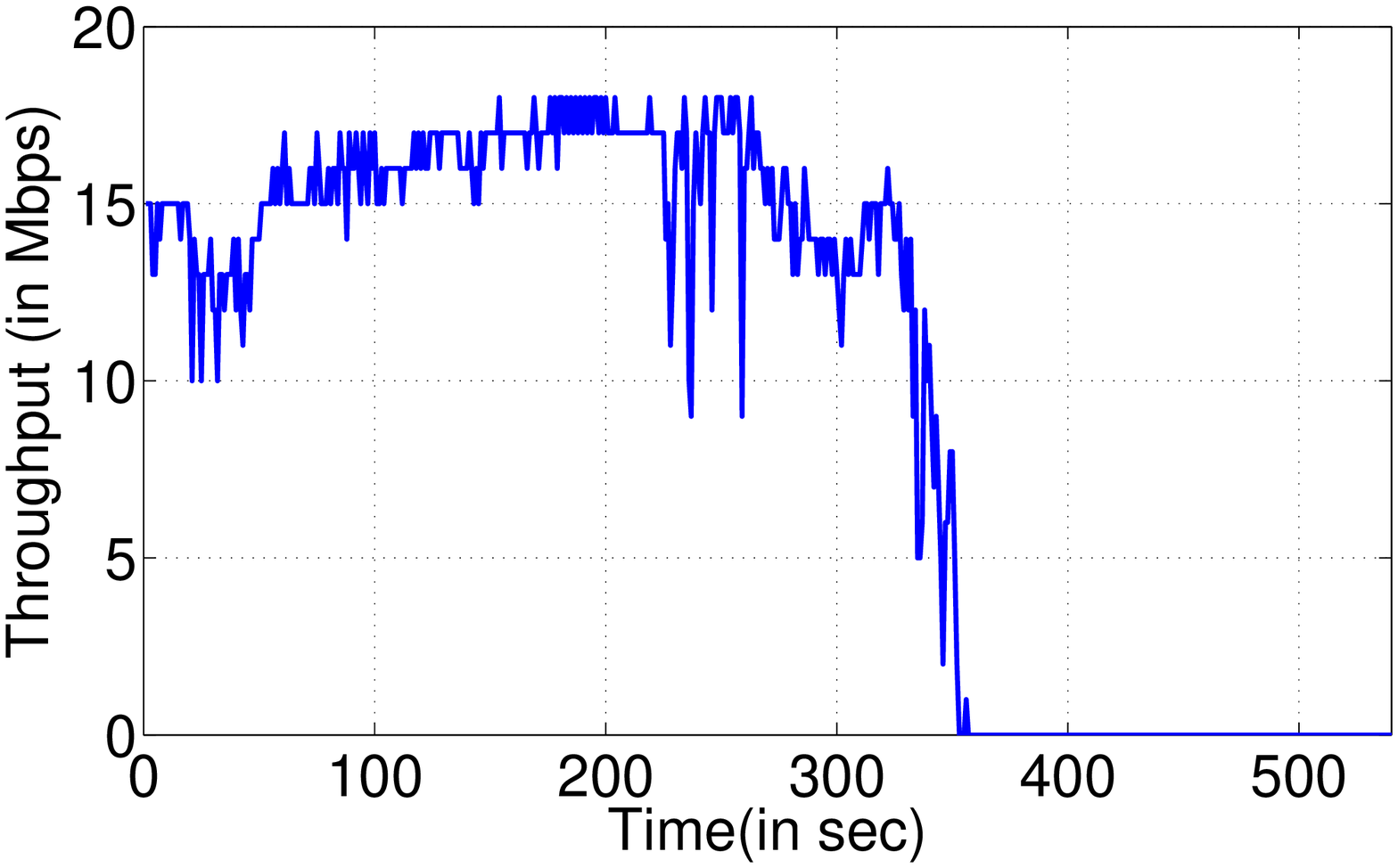}
\label{fig:iperf_walk}
} \hspace{-4mm}
\subfigure[]{
\includegraphics[width=0.33\linewidth]{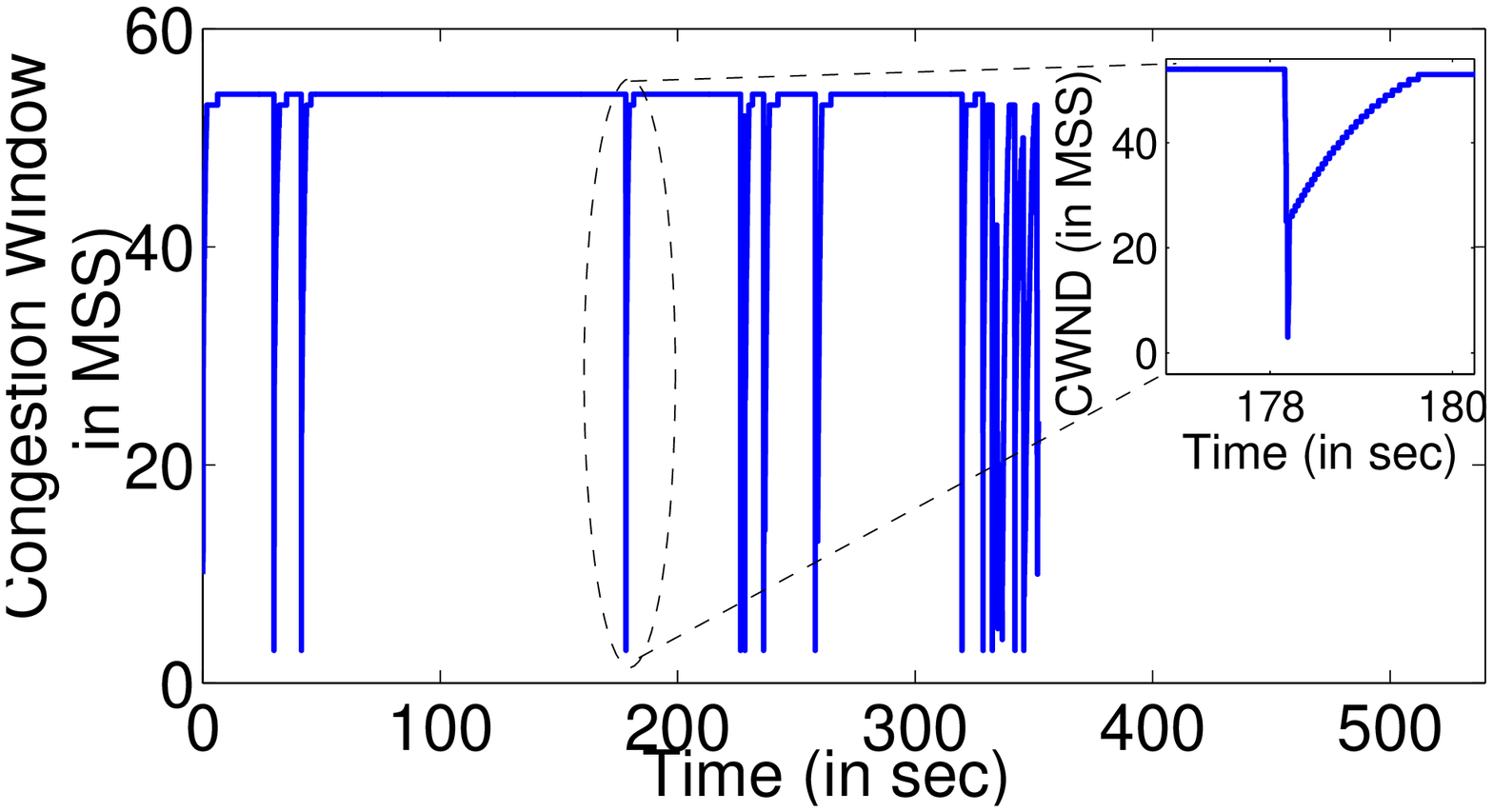}
\label{fig:cwnd_walk}
}\hspace{-4mm}
\subfigure[]{
\includegraphics[width=0.33\linewidth]{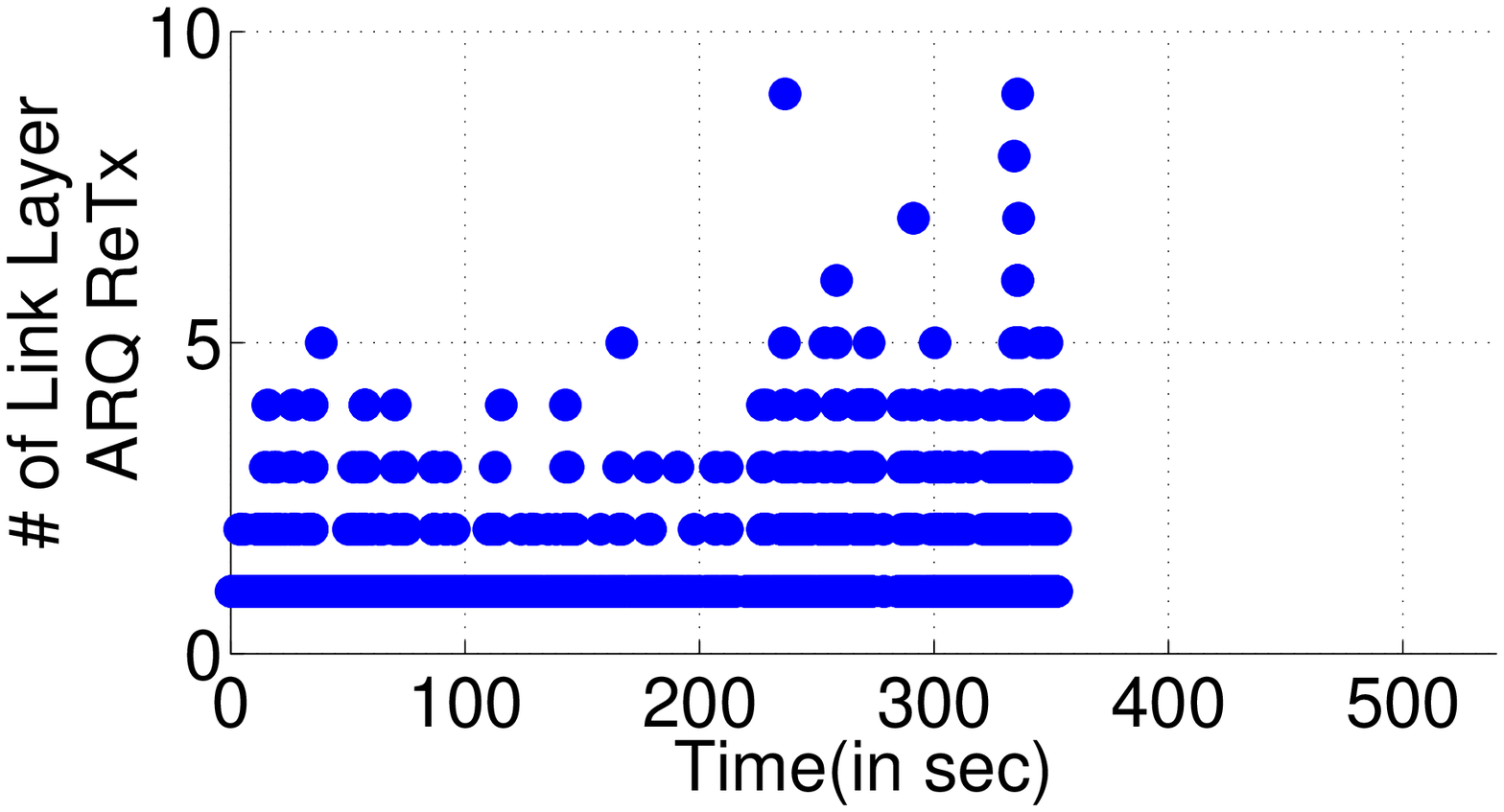}
\label{fig:mac_retx_count_walk}
}\hspace{-1cm}

\subfigure[]{
\includegraphics[width=0.34\linewidth]{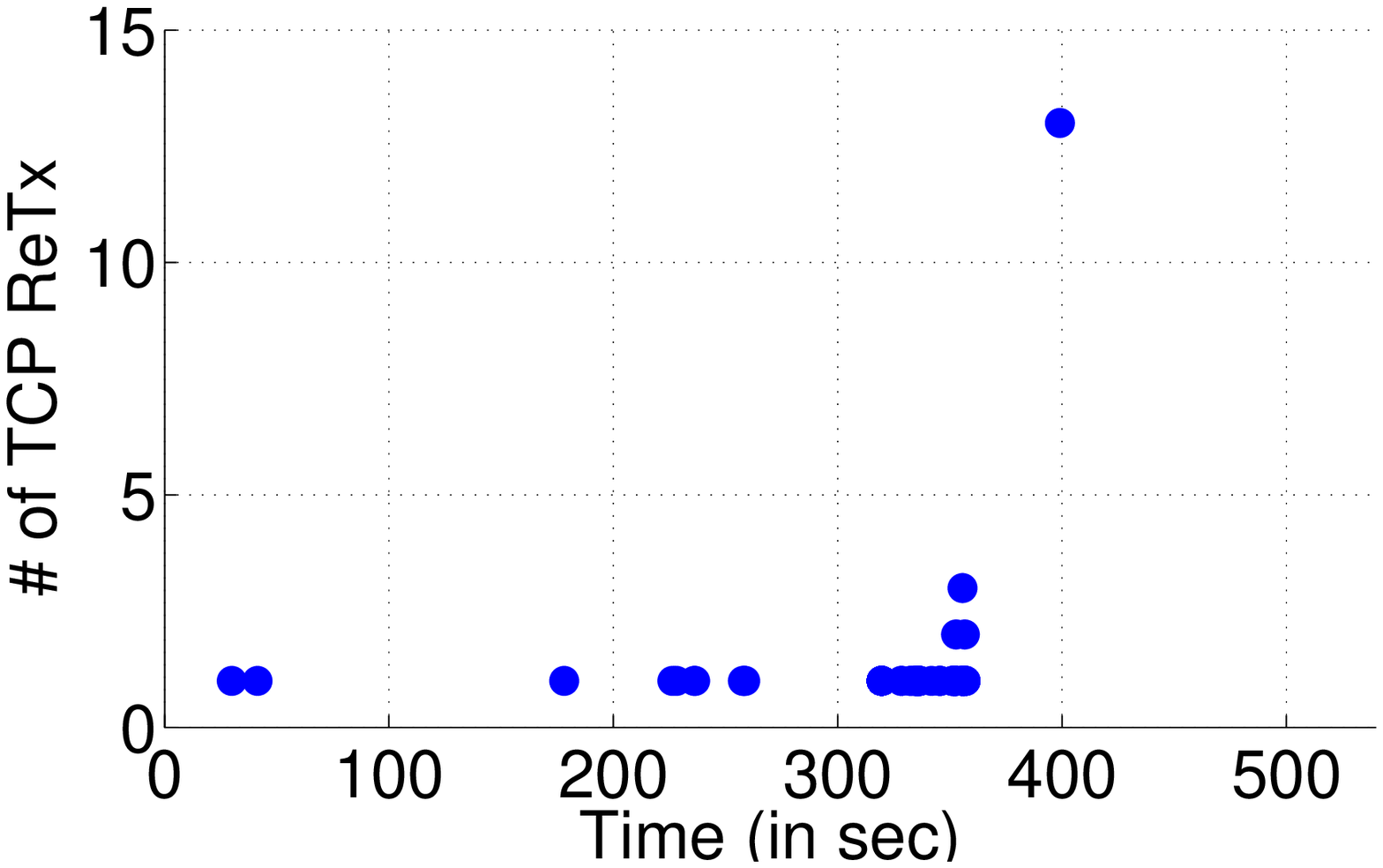}
\label{fig:tcp_retx_count_walk}
}
\hspace{-5mm}
\subfigure[]{
\includegraphics[width=0.33\linewidth]{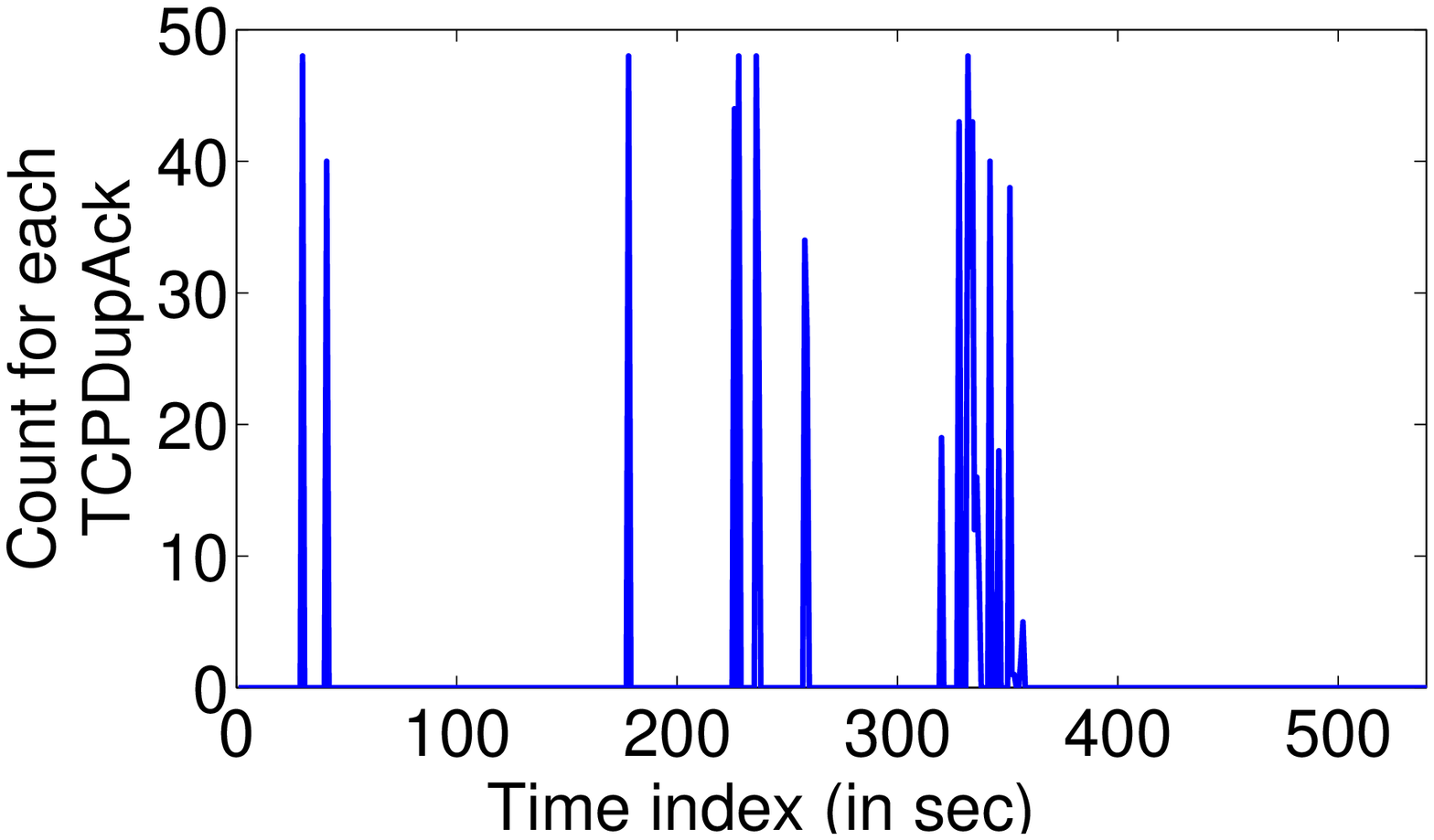}
\label{fig:tcp_dup_ack_count_walk}
}
\hspace{-5mm}
\subfigure[]{
\includegraphics[width=0.33\linewidth]{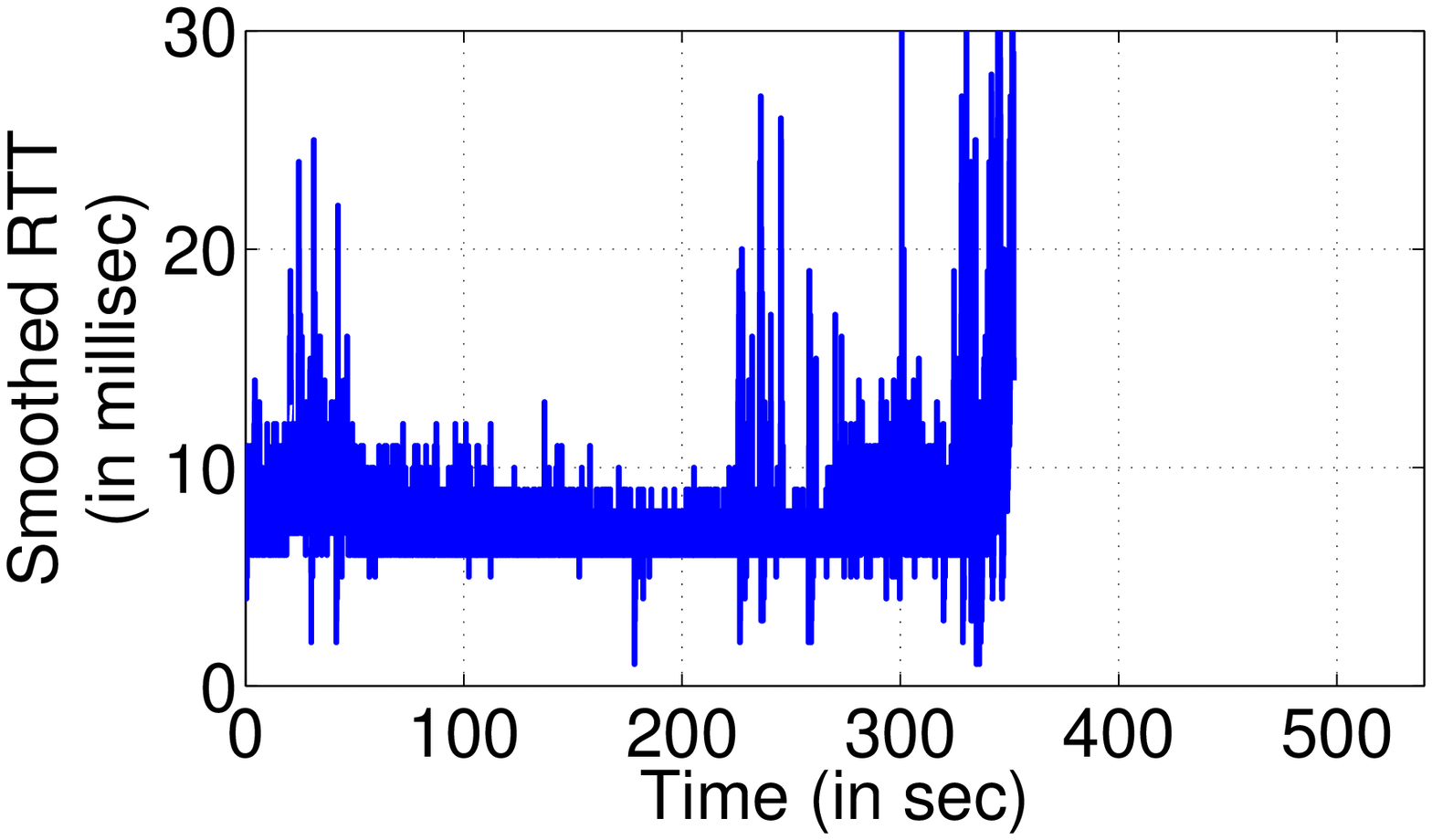}
\label{fig:srtt_walk}
}\hspace{-3cm}
\label{fig:results_walk}
\caption{Experimental results for pedestrian mobility scenario: \protect \subref{fig:iperf_walk} Iperf throughput in Mbps when the receiver is moving along Route $1$ (in Fig.~\ref{fig:exp_drive}); \protect \subref{fig:cwnd_walk} Congestion window trajectory with a callout (upper right) highlighting the fall and slow-start; \protect \subref{fig:mac_retx_count_walk} Number of ARQ retransmissions on WiFi link for each retransmitted ARQ frame, showing increased ARQ retries as the receiver moves away from the line-of-sight of the AP; \protect \subref{fig:tcp_retx_count_walk} Number of TCP retransmissions   for each retransmitted TCP segment; \protect \subref{fig:tcp_dup_ack_count_walk} Number of DupAcks received for individual TCP packet; \protect \subref{fig:srtt_walk} Smoothed RTT computed  by the sender for each Ack reception}.
\vspace{-0.5cm}
\end{figure*}

In order to emulate a pedestrian walking, the receiver moves at an average speed of
$6$ mph on the route A-B-C-D (Route $1$, as highlighted in Fig~\ref{fig:exp_drive}). 
In Fig.~\ref{fig:iperf_walk}, we see the throughput 
increases as the receiver moves closer to the AP, then drops as it moves away.
 The connection finally breaks at around $350$ seconds, when it moves out of line-of-sight of the AP. 
Between $0$ and $100$s, the congestion window, \textit{cwnd}, suffers $2$ sudden drops as seen in Fig~\ref{fig:cwnd_walk}, e.g., at around $30$ and $40$ seconds, respectively. 
\textit{cwnd}'s drop adversely affects the throughput (see Fig.~\ref{fig:iperf_walk}).
To trace the cause of window drop, note that from Fig.~\ref{fig:mac_retx_count_walk}, we see there are more ARQ retransmissions around the times the window drops, compared to times when the window (and
throughput) is high (e.g., at around time $100$s). Moreover, in spite of the link layer's frequent ARQ attempts, end-to-end losses are still seen by TCP sender, as indicated by the two bursts of duplicate acknowledgements (DupAck) from the receiver, shown in Fig.~\ref{fig:tcp_dup_ack_count_walk}.
These occur around the time instants when \textit{cwnd} drops. Specifically, the number of DupAcks at 
time $30$s and $40$s are $48$ and $40$, respectively. TCP Reno governs the sender's reaction to DupAcks:
 when it sees three DupAcks, a fast retransmission of the lost packet takes place (seen in Fig.~\ref{fig:tcp_retx_count_walk}).
In this simple topology (sender to AP over a GigE link, AP to receiver is a 802.11g WiFi link), the retransmitted frame from the sender node has to wait in the transmit queue of the AP, until all the buffered frames at the AP are delivered. Each of those frames draining out of the queue to the
receiver on the WiFi link again result in an additional DupAck being sent upstream. Because the sender
has transmitted as many packets as the outstanding window allowed, during this period
when the WiFi link drains the queue, the sender essentially has stopped sending and waits until the new acknowledgement is received. Unlike the specification on the fast retransmit, the typical Linux implementation of TCP reduces the window (\textit{cwnd}) first by half on receiving the $3$ DupAcks, and then continues to
reduce \textit{cwnd} further by one for every $2$ additional DupAcks (thus preventing transmission of new data, as we 
observe from the pcap traces).
When the retransmitted packet is received, its Ack, acknowledging all the outstanding data
will enable the sender to begin sending packets. But with the \textit{cwnd}
`deflation' rule, from~\cite{rfc2581} that is meant to
avoid a burst of data being sent into the network due to sudden growth of the congestion window, 
the sender reduces \textit{cwnd} down to 1 and performs a slow start (by the implementation in Linux again,
a more conservative choice than recommended in ~\cite{ rfc2582}). This causes a significant throughput
penalty as the window then recovers through slow start. We will look at this window recovery process again in the next subsection.
While some of this behavior may be attributed to the phenomenon of `buffer bloat' ~\cite{get-bufferbloat}, we observe
that the buffering of $50$-$60$ packets at the AP is not unusual when it is the `bridge' between the GigE link on one side and a 
WiFi link with highly variable bandwidth.

\vspace{-1mm}
\subsection{A Slow-Drive Scenario}
\vspace{-1mm}
In this experiment, we performed a set of experiments with the receiver in a car
driven at a speed of 10mph along E-F-G-H-D (Route $2$ in Fig.~\ref{fig:exp_drive}), 
emulating a car going through a traffic intersection. We also
chose this route in order to analyze the consequences of losing line-of-sight from the AP during an ongoing I2V transmission. 
The throughput \ref{fig:iperf_parking_lot}, initially starts out by increasing quickly
and being stable, i.e. staying around $17$Mbps from $0$s to $20$s. 
As the car moves further into the parking lot and the line of sight to the AP begins to be blocked by the building, throughput drops significantly. 
Fig.~\ref{fig:mac_retx_count_parking_lot} shows that ARQ retransmissions in this interval
are relatively sparse and the retransmit count for each link layer frame does not exceed 1 prior to $20$s; 
subsequently, the number of ARQ retransmits increases substantially. 
As a consequence, the TCP layer observes residual loss, and at a time
just after $t=40$s, two packets are lost.
From $t=30$s to $50$s, multiple DupAcks are received at the sender (see Fig.~\ref{fig:tcp_dup_ack_count_parking_lot}), causing the sender to reduce the congestion window size. This occurs multiple
times as TCP packets are lost in quick succession (and the \textit{cwnd} has
not built up to the larger value (see Fig.~\ref{fig:cwnd_parking_lot}).
The sending TCP uses fast retransmit in response to the DupAcks
(see Fig.~\ref{fig:seq_tcp_retx_parking_lot}), but because of the burst of DupAcks the
\textit{cwnd} value drops down to 1 anyway. Starting from time = $50$s, the
sender is not able to receive any acknowledgements from the receiver
(loss of line-of-sight), and the TCP connection does not make progress.
The sender experiences
repeated TCP timeouts and retransmits the packets unsuccessfully, as seen in Fig.~\ref{fig:seq_tcp_retx_parking_lot}. During this interval, the retransmission timeout value grows exponentially. 
Eventually, when the vehicle moves back within sight of the AP, the TCP connection (which
did not drop during this nearly 40-second period) continues and data transfer throughput
begins to build back up (with the congestion window building up) after the sender receives the expected Ack at around $88$s.  Note though that there is one more episode of a residual loss resulting in
a cascade of DupAcks, reduction in \textit{cwnd} and a `hit' to the throughput around $90$s as the vehicle 
continues to move.
\subsubsection{Some losses are worse than others}
\begin{figure*}[!t]
\centering
\hspace{-1cm}
\subfigure[]{
\includegraphics[width= 0.34\linewidth]{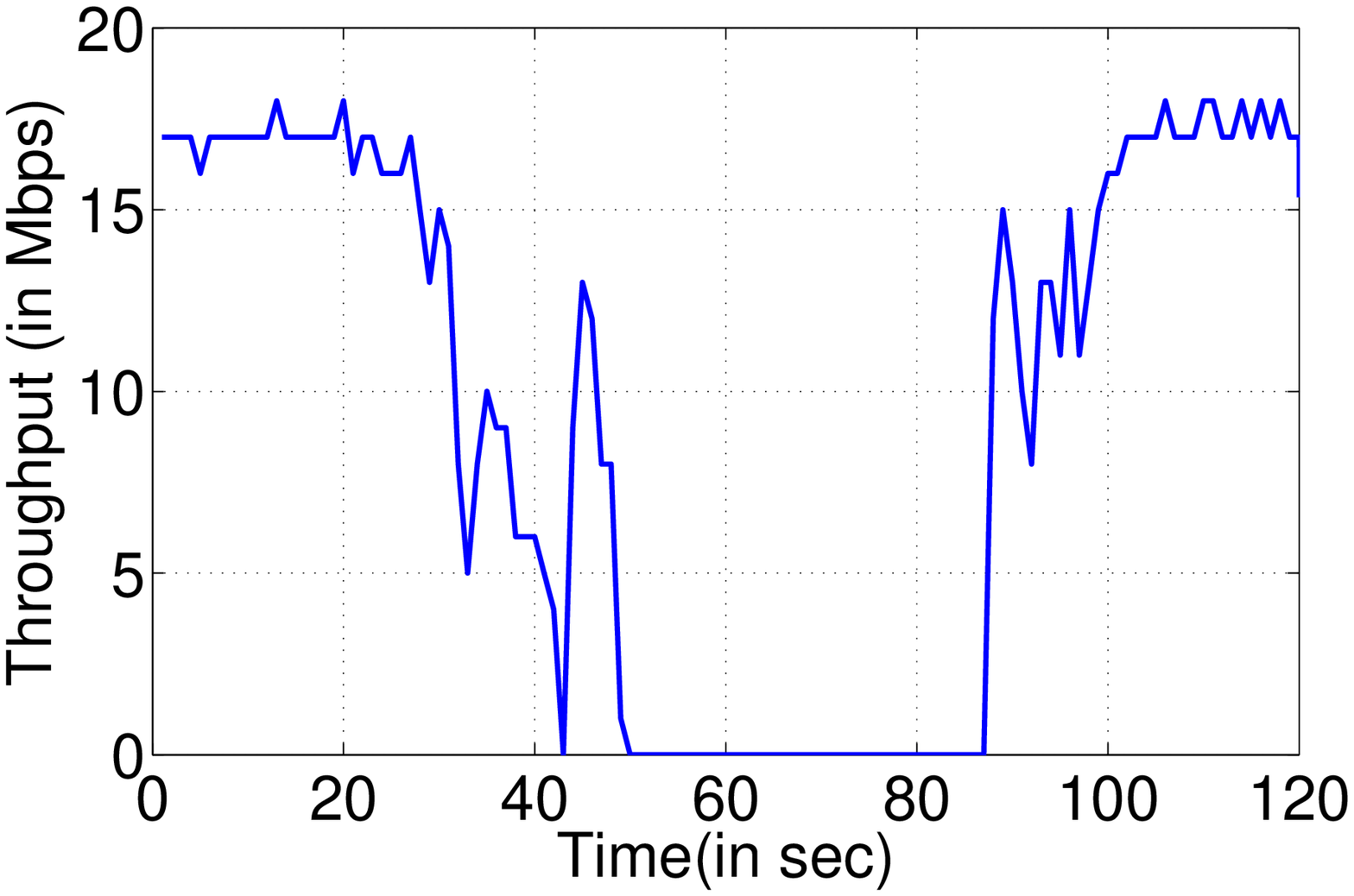}
\label{fig:iperf_parking_lot}
}
\hspace{-6mm}
\subfigure[]{
\includegraphics[width=0.33\linewidth]{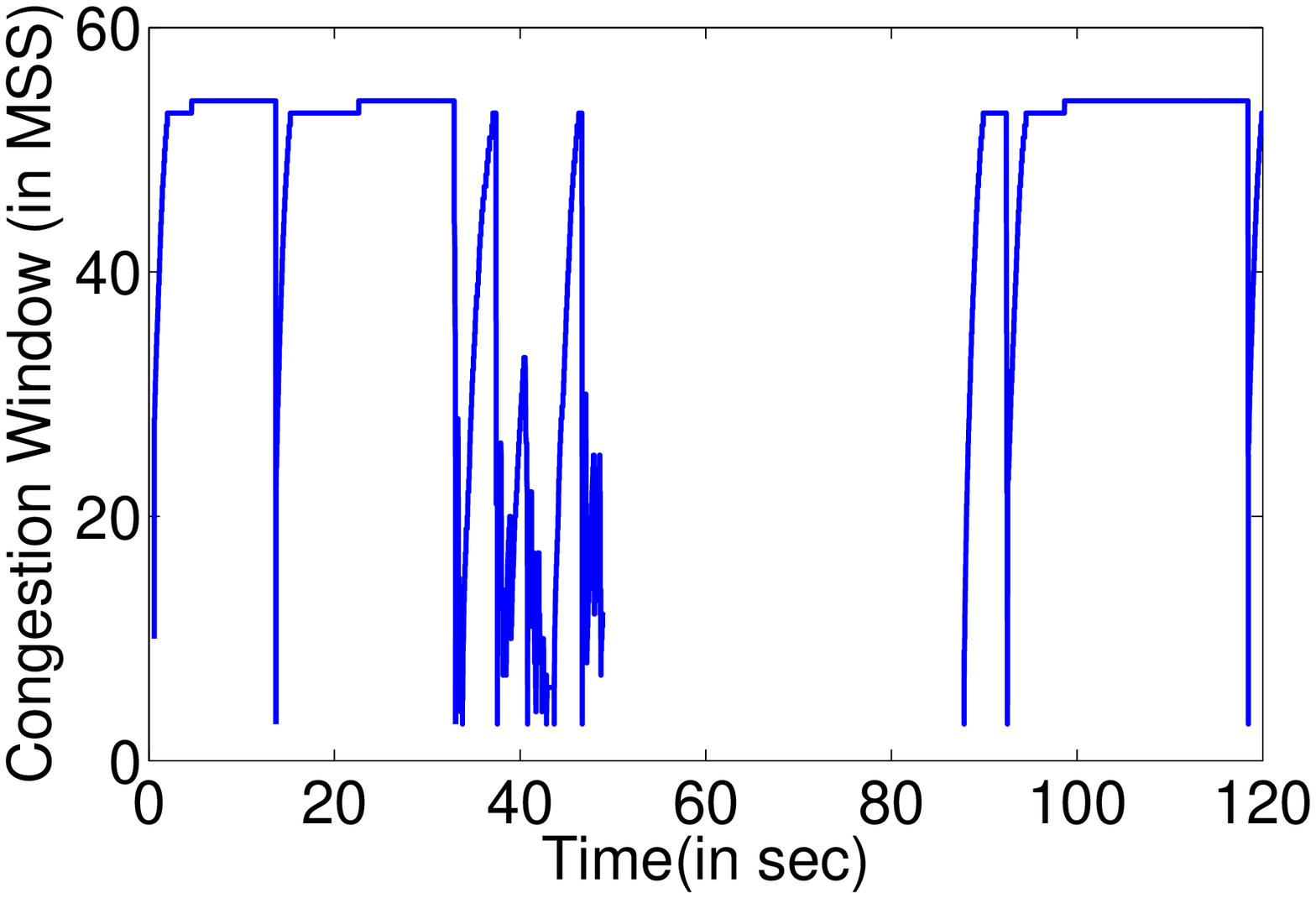}
\label{fig:cwnd_parking_lot}
}
\hspace{-6mm}
\subfigure[]{
\includegraphics[width=0.34\linewidth]{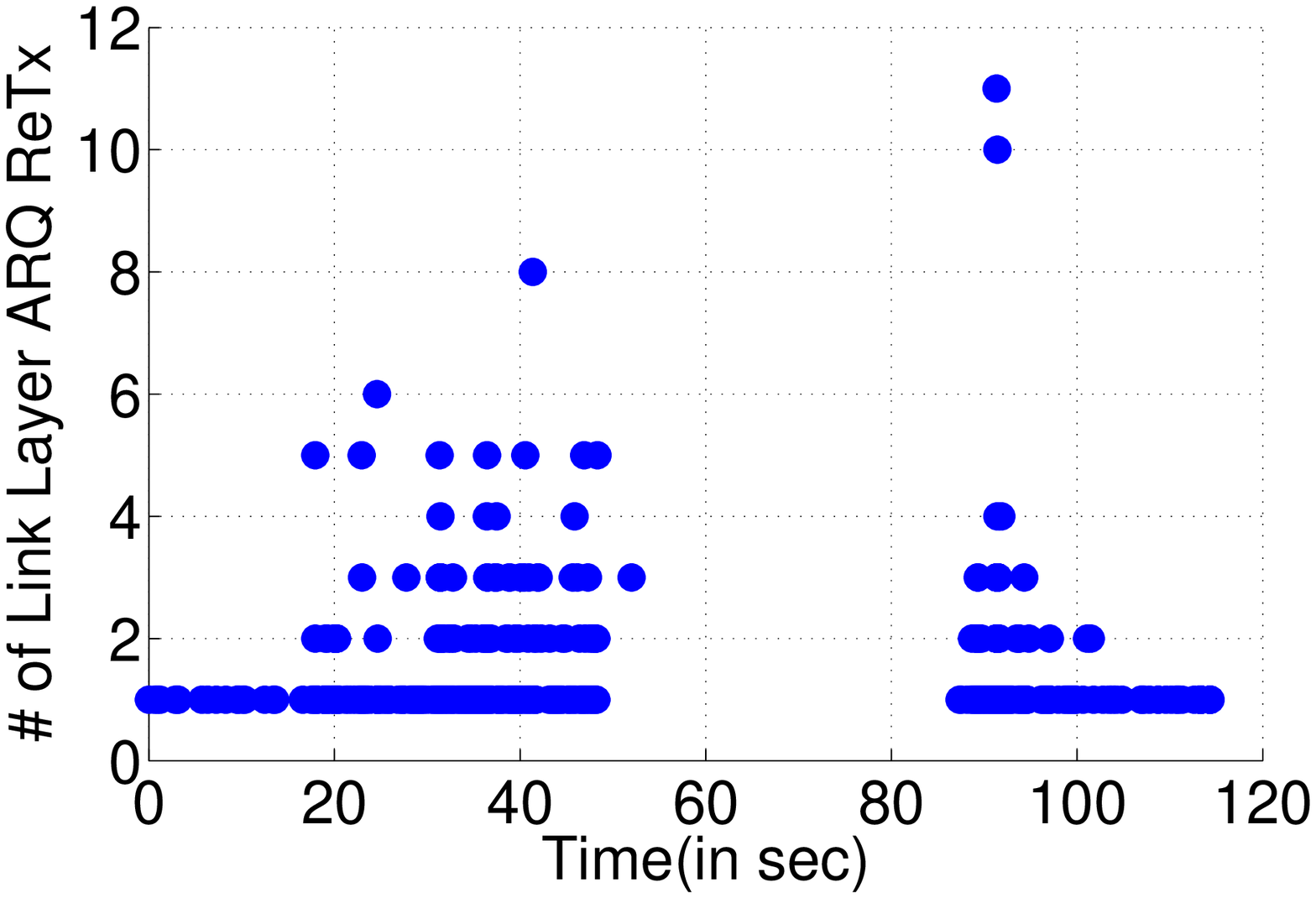}
\label{fig:mac_retx_count_parking_lot}
}
\hspace{-1cm}
\subfigure[]{
\includegraphics[width=0.35\linewidth]{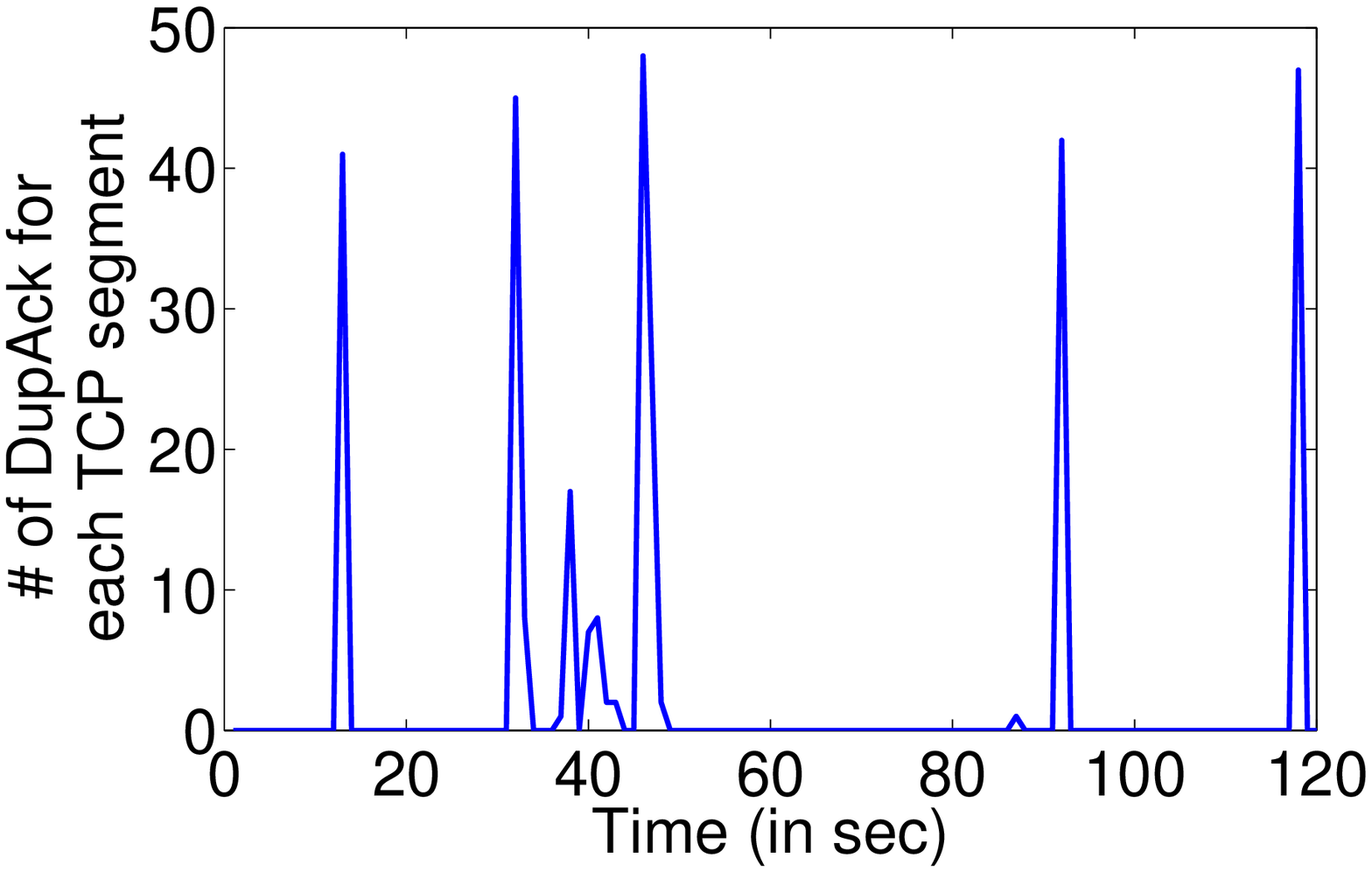}
\label{fig:tcp_dup_ack_count_parking_lot}
}
\hspace{-8mm}
\subfigure[]{
\includegraphics[width=0.33\linewidth]{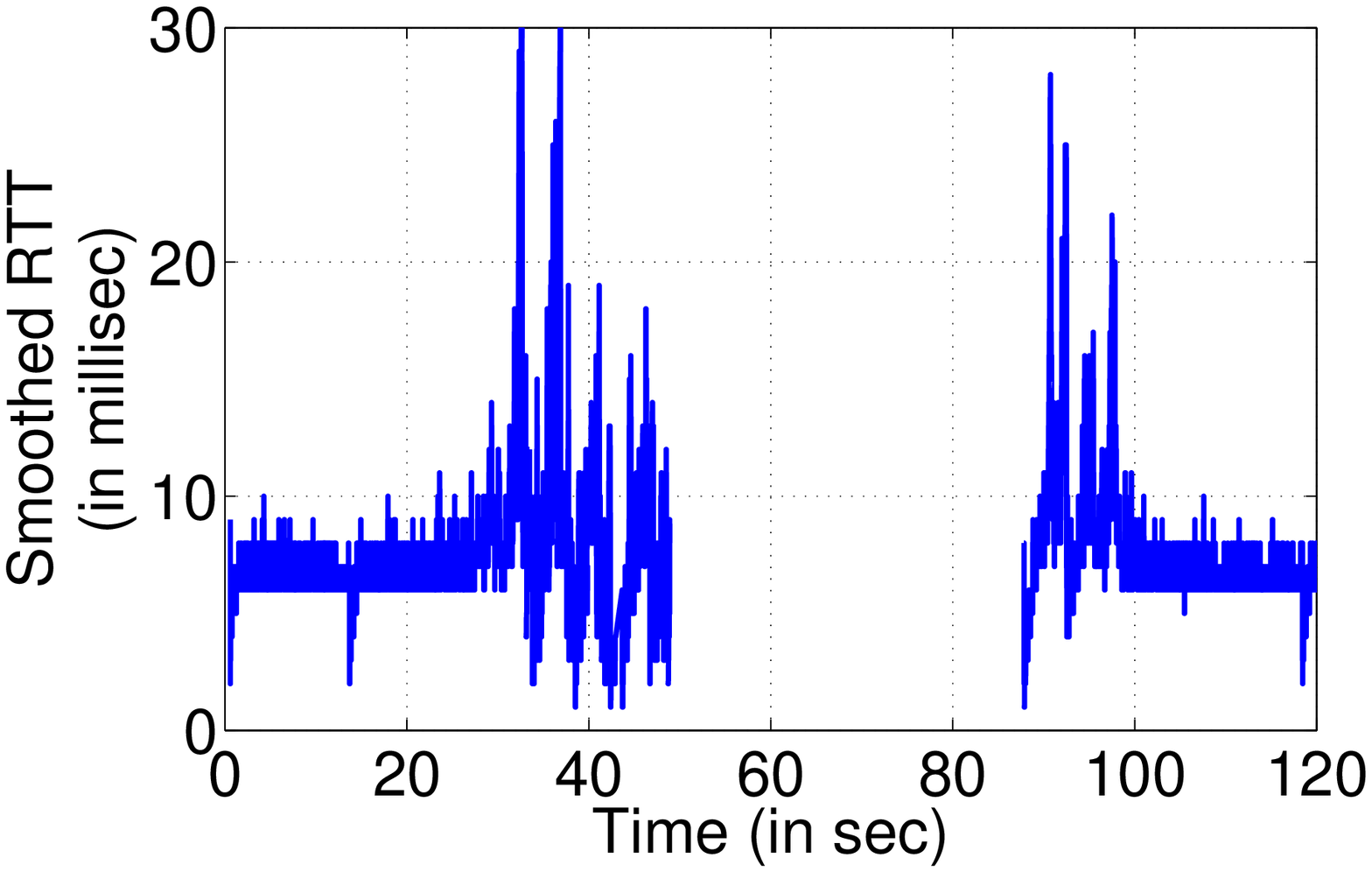}
\label{fig:srtt_parking_lot}
}\hspace{-8mm}
\subfigure[]{
\includegraphics[width=0.34\linewidth]{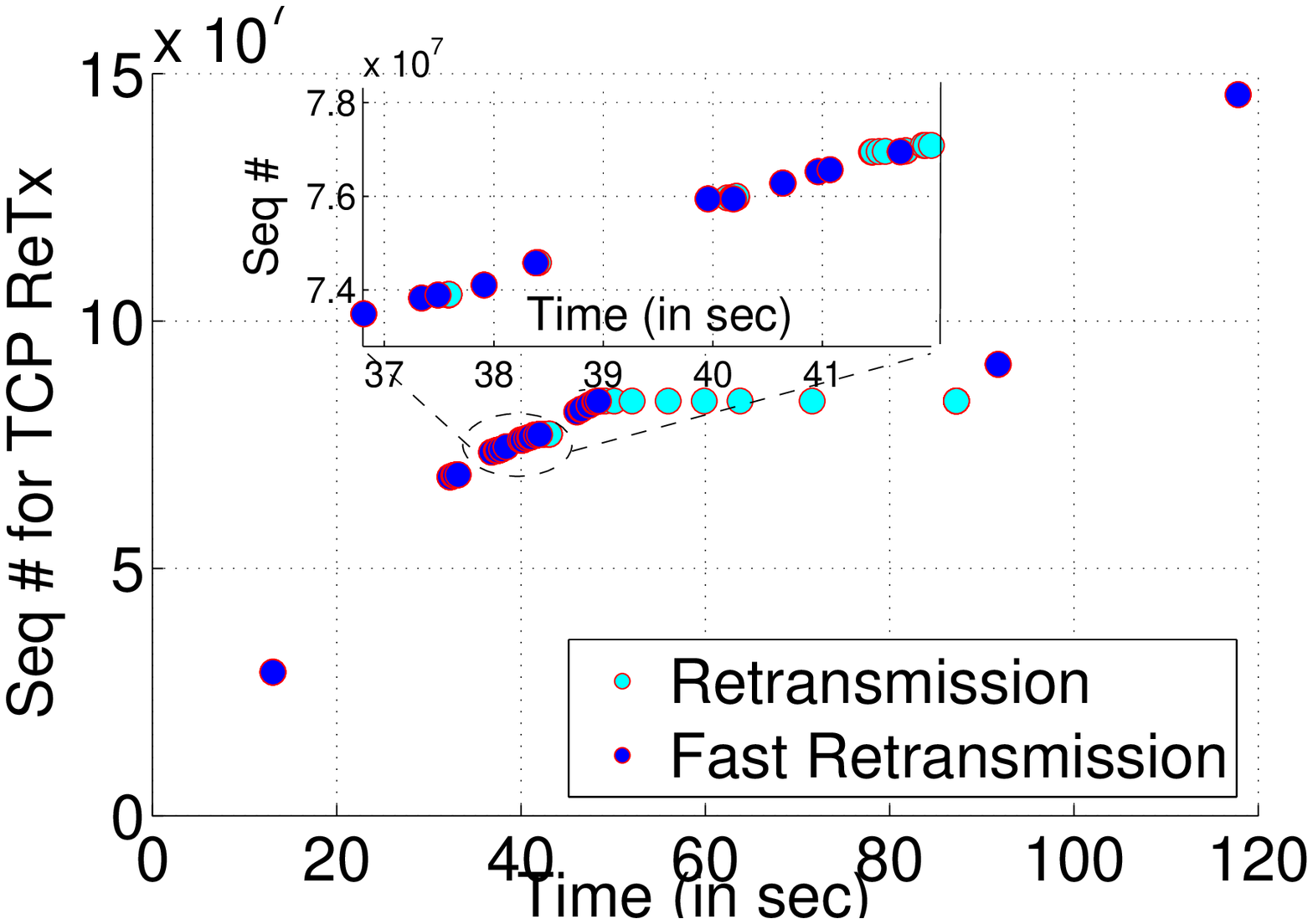}
\label{fig:seq_tcp_retx_parking_lot}
}
\hspace{-1cm}
\label{fig:results_parking_lot}
\caption{Experimental results for a slow drive scenario along Route $2$ (in Fig.~\ref{fig:exp_drive}): \protect \subref{fig:iperf_parking_lot} Iperf throughput in Mbps, showing the throughput going down to zero when the vehicle moves out of the line-of-sight of the AP and later coming back up as it moves into view; \protect \subref{fig:cwnd_parking_lot} Congestion window trajectory from tcp probe. \textit{cwnd} is logged for every Ack received, and no Acks were received from $48$ to $87$ seconds;  \protect \subref{fig:mac_retx_count_parking_lot} Number of ARQ retransmissions  on the WiFi link, showing that the AP stops retransmitting in-between due to loss association; \protect \subref{fig:tcp_dup_ack_count_parking_lot}  Number of DupAcks received for individual TCP packet; \protect \subref{fig:srtt_parking_lot} Smoothed RTT computed by the sender for each Ack reception. The plot is discontinuous when the AP loses association; \protect \subref{fig:seq_tcp_retx_parking_lot} Sequence number of each retransmitted TCP segment, showing that the TCP connection is not lost. The callout (on top) shows several segments getting retransmitted in a short amount of time, both due to fast retransmits as well as RTOs. The number of TCP retransmissions for each segment is not included for brevity, but its behavior is similar to Fig.~\ref{fig:tcp_retx_count_walk}}
\vspace{-0.6cm}
\end{figure*}

In the two experiments we described so far, TCP sees packet loss and seeks to
recover through fast retransmission on receiving three DupAcks, but the
reception of a cascade of DupAcks results in the sender's congestion window dropping all the way down to 1. However, we observe that some of the losses are more
harmful, in terms of impacting throughput than others.
In the `slow-drive' scenario, the losses causing the \textit{cwnd} to drop at around $t=32$s and $t=92$s (shown in Fig.~\ref{fig:cwnd_parking_lot}), result in 
a throughput reduction of about $10$Mbps and $7$Mbps (see Fig.~\ref{fig:iperf_parking_lot}).
On the other hand, the window drop at $t=13$s, does not cause a substantial throughput degradation. We call these two types of loss `penalizing loss' and `moderate loss' respectively. In fact, the determining factor of whether a loss event hurts throughput or not appears to be the current RTT experienced by the TCP 
connection. 

We plot the Smoothed RTT (SRTT) \footnote{SRTT is the sender's smoothed estimate of RTT, a reflection of instantaneous link quality (loss, delay) on the
end-end connection and influences TCP's window growth and in computing the retransmission timeout value.}
estimated at the sender when receiving each Ack packet. 
From Fig.~\ref{fig:srtt_parking_lot} we observe that at $t=13$s, when the sender enters the slow start phase to rebuild the window after it `crashes' to 1, 
the SRTT value is approximately $2$ms, and only grows
a little, without  exceeding $10$ms, till $t=20$s. Further, the time it
takes to build the congestion window back up is only $1.57$ seconds, and the
throughput does not fall significantly.
This is the `moderate loss' case.
In contrast, SRTT increases dramatically starting at $92$s.
At this point, the loss causes the cascade of DupAcks,
the \textit{cwnd} value crashes, and the throughput also falls. Just before
this we observe that the number of link layer ARQ is also higher, with
some packets being retransmitted 10 or 11 times. The SRTT grows to a
much higher value ($25$ms). This in turn slows down the congestion window growth that now takes $2.11$ seconds to build back up. This penalizes the throughput achieved - hence the term `penalizing loss'. 
We observe in all the scenarios where the loss occurs with a large SRTT,
the hit in the throughput for such a penalizing loss is noticeable and
influential in impacting application performance.
For a static scenario (described below), we observe that the current conservative window recovery process does not penalize the throughput from the
application's point of view. However, in I2V or even in
pedestrian mobility scenarios, where the the user has 
only a short time of association with the WiFi channel, recovering from a 
`penalizing loss' is detrimental to application performance. Our current work
is to find improvements to avoid such situations.

\vspace{-1mm}
\subsection{Static Scenario}
\begin{figure*}[!t]
\label{fig:results_static}
\centering
\hspace{-1cm}
\subfigure[]{
\includegraphics[width= 0.3\linewidth]{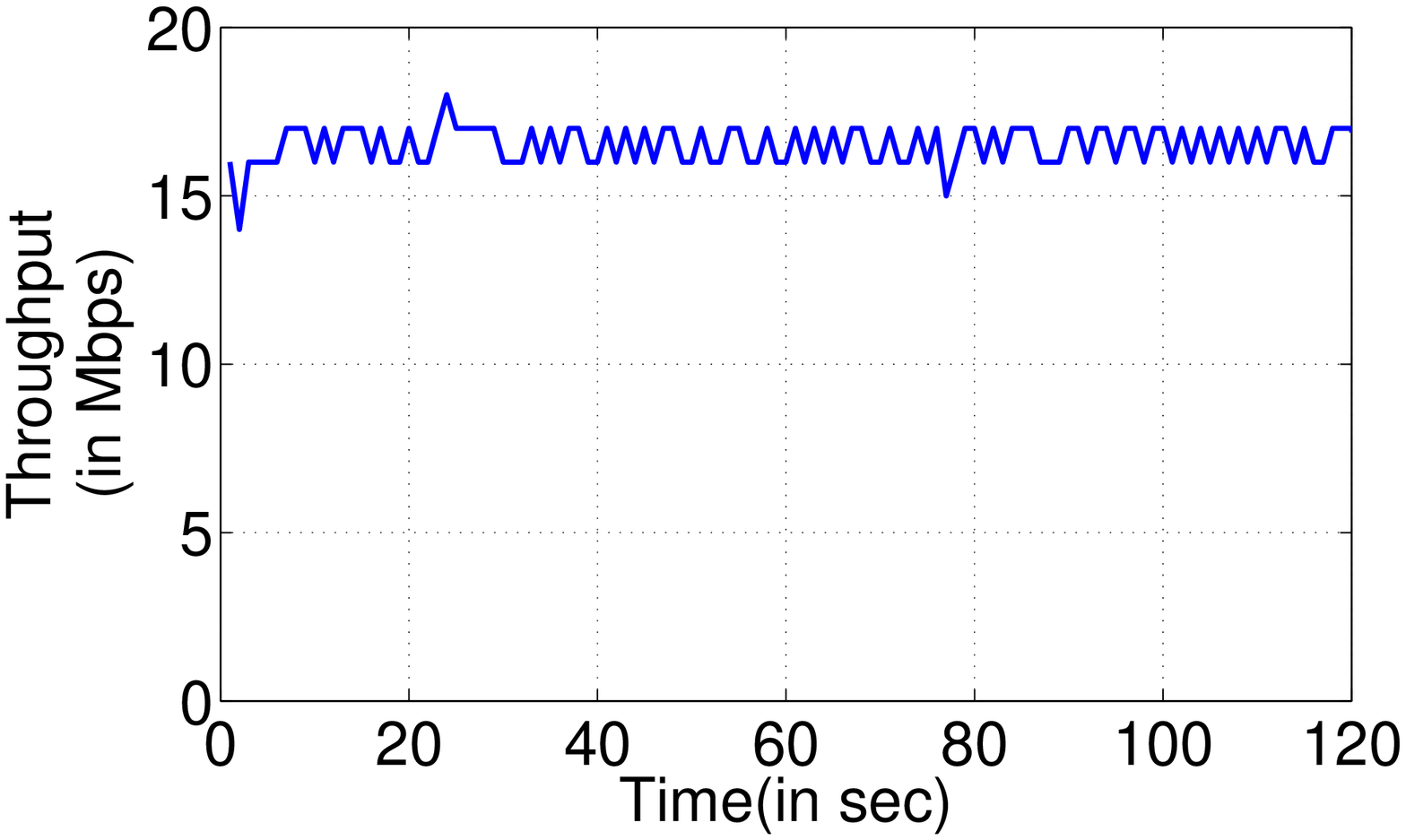}
\label{fig:iperf_static}
}
\hspace{-6mm}
\subfigure[]{
\includegraphics[width=0.3\linewidth]{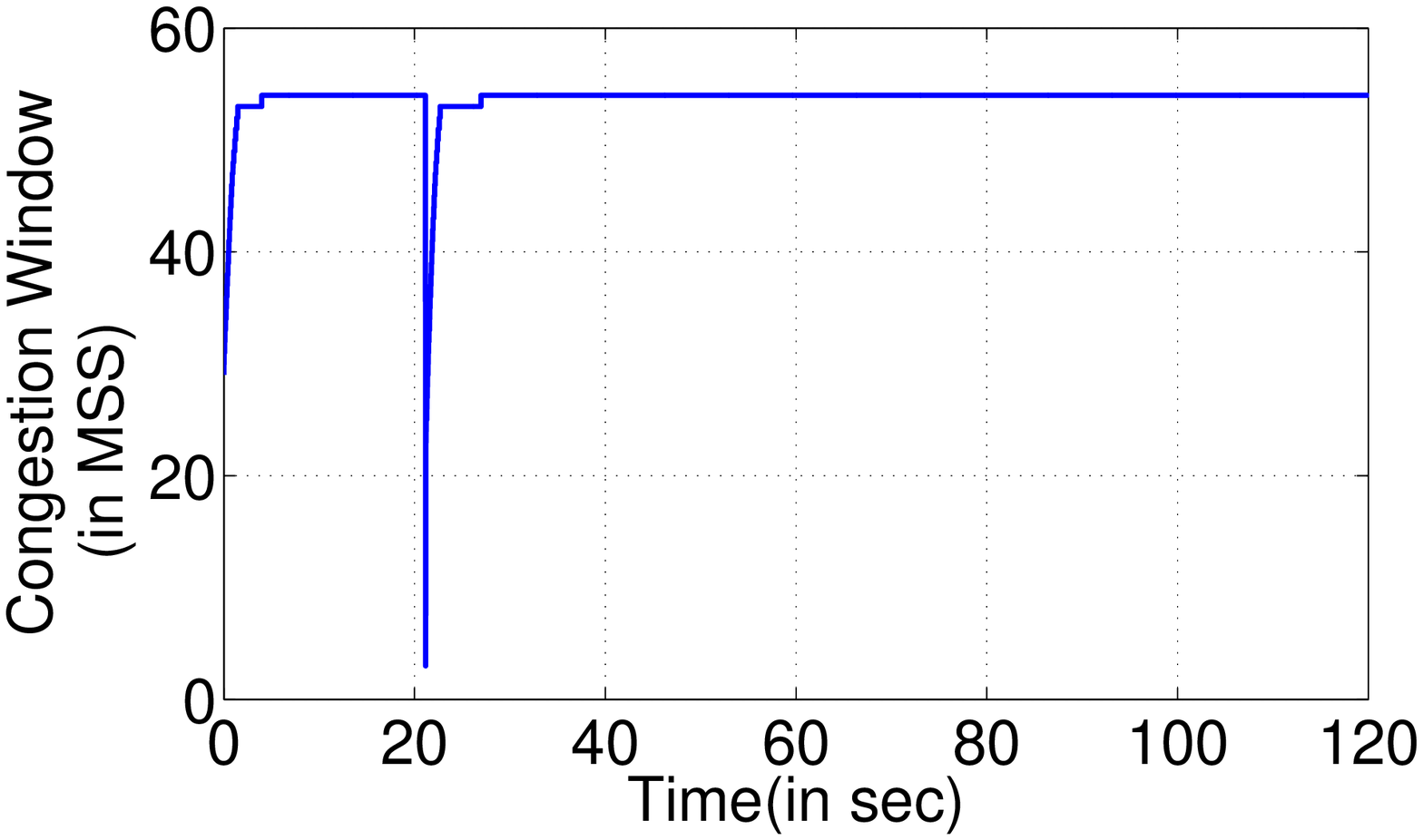}
\label{fig:cwnd_static}
}
\hspace{-6mm}
\subfigure[]{
\includegraphics[width=0.3\linewidth]{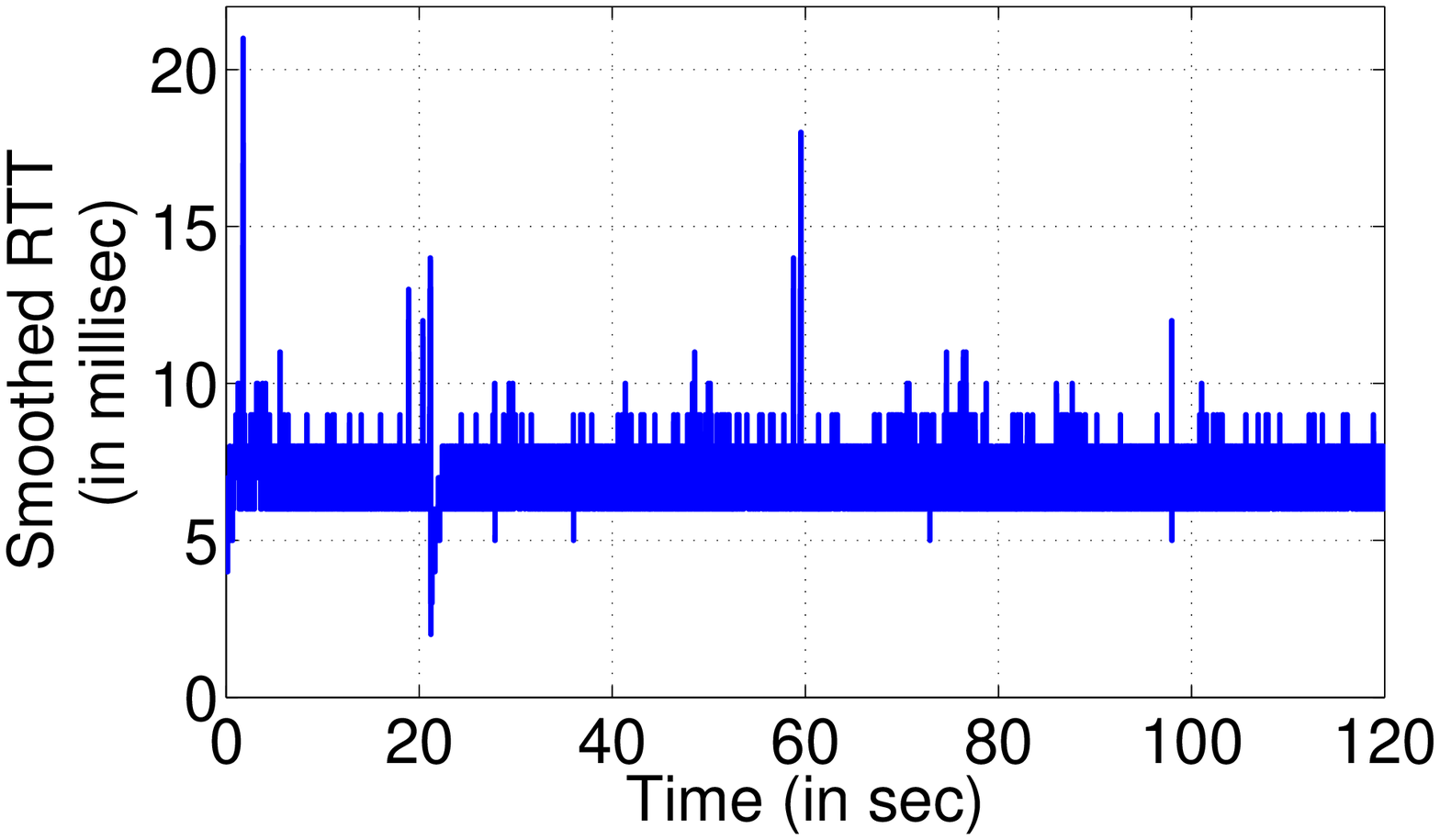}
\label{fig:srtt_static}
}
\hspace{-1.2cm}
\caption{Experimental results for a static user scenario, with the receiver fixed at Point `B' in Fig.~\ref{fig:exp_drive}: \protect \subref{fig:iperf_static} Iperf throughput in Mbps showing stable throughput in spite of \textit{cwnd} drop; \protect \subref{fig:cwnd_static} Congestion window trajectory; \protect \subref{fig:srtt_static} Smoothed RTT computed by the sender for each Ack reception}
\end{figure*}

\vspace{-1mm}
To further understand if movement is the cause of `penalizing loss',
we conducted the same experiment with the car parked $52.52$m away from the AP 
(point B in Fig.~\ref{fig:exp_drive}).
Results of this experiment are shown in Figs~\ref{fig:iperf_static} -~\ref{fig:srtt_static}. From the \textit{cwnd} plot, we can see that only one TCP loss event occurs during the course of this experiment. When the \textit{cwnd} starts to recover after dropping down to one (at around $22$s), SRTT grows moderately, not exceeding $8$ms. This suggests that when the client is static, the moderate
loss event does not penalize the throughput as much. We hope to look at the 
precise sequence of events to understand the root cause for these different
loss situations, as we develop solutions to improve 
content delivery performance.

%
%

\section{Summary}

We experimentally studied the delivery of large volume
content to vehicular clients over a 802.11 WiFi infrastructure, using TCP.
TCP's reaction to packet loss over the WiFi link can be particularly severe,
thereby substantially reducing the throughput achieved, even
for a single packet loss. Since a vehicular 
client may be associated with an AP for a short period, it is important
to not suffer this throughput penalty. 

Our results suggest that I2V communication over WiFi has the potential to support content delivery to both 
pedestrians and vehicles. However, care needs to be taken. First, our experiments suggest that the APs should be judiciously placed, where the vehicular connection time is extended as much as possible. 
For example, they could be at intersections where vehicles 
often stop and wait, especially if placed at a higher elevation to retain line-of-sight as far as possible. 
Moreover, the 802.11 MAC  and the TCP transport layers could be integrated more intelligently. For example, using FEC or network coding, TCP can be sequence number agnostic and not be `stuck' on the loss of a particular packet.  
Possible network layer solutions include content caching and storage at the routers in conjunction with mobility-prediction and multihoming via cellular and WiFi. Many of these ideas are explicit design features of clean-slate Future Internet architectures such as MobilityFirst~\cite{raychaudhuri2012mobilityfirst}. 
It is also becoming increasingly clear that the transport layer needs to adapt so as to 
not over-react to non-congestion losses, as these occur frequently enough in our experiments, despite the use of link layer ARQ.

\vspace{-3mm}





\bibliographystyle{IEEEtran}
\bibliography{library}

\begin{thebibliography}{10}
\providecommand{\url}[1]{#1}
\csname url@samestyle\endcsname
\providecommand{\newblock}{\relax}
\providecommand{\bibinfo}[2]{#2}
\providecommand{\BIBentrySTDinterwordspacing}{\spaceskip=0pt\relax}
\providecommand{\BIBentryALTinterwordstretchfactor}{4}
\providecommand{\BIBentryALTinterwordspacing}{\spaceskip=\fontdimen2\font plus
\BIBentryALTinterwordstretchfactor\fontdimen3\font minus
  \fontdimen4\font\relax}
\providecommand{\BIBforeignlanguage}[2]{{%
\expandafter\ifx\csname l@#1\endcsname\relax
\typeout{** WARNING: IEEEtran.bst: No hyphenation pattern has been}%
\typeout{** loaded for the language `#1'. Using the pattern for}%
\typeout{** the default language instead.}%
\else
\language=\csname l@#1\endcsname
\fi
#2}}
\providecommand{\BIBdecl}{\relax}
\BIBdecl

\bibitem{safety-url}
\BIBentryALTinterwordspacing
``Vehicle-to-infrastructure (v2i) communications for safety.'' [Online].
  Available: \url{http://www.its.dot.gov/factsheets/v2isafety_factsheet.htm}
\BIBentrySTDinterwordspacing

\bibitem{2001-its}
L.~Figueiredo, I.~Jesus, J.~Machado, J.~Ferreira, and J.~M. de~Carvalho,
  ``Towards the development of intelligent transportation systems,'' in
  \emph{Intelligent Transportation Systems}, 2001.

\bibitem{muni-map}
\BIBentryALTinterwordspacing
``Municipal wireless network projects map.'' [Online]. Available:
  \url{http://news.cnet.com/Municipal-broadband-and-wireless-projects-map/2009-1034_3-5690287.html}
\BIBentrySTDinterwordspacing

\bibitem{google-wifi}
\BIBentryALTinterwordspacing
``Google wifi.'' [Online]. Available:
  \url{http://en.wikipedia.org/wiki/Google_WiFi}
\BIBentrySTDinterwordspacing

\bibitem{des-imc-10}
\BIBentryALTinterwordspacing
P.~Deshpande, X.~Hou, and S.~R. Das, ``Performance comparison of 3g and
  metro-scale wifi for vehicular network access,'' in \emph{Proc. of the 10th
  ACM SIGCOMM Conference on Internet Measurement}, New York, NY, USA, 2010.
  [Online]. Available: \url{http://doi.acm.org/10.1145/1879141.1879180}
\BIBentrySTDinterwordspacing

\bibitem{ott2004drive}
J.~Ott and D.~Kutscher, ``Drive-thru internet: Ieee 802.11 b for" automobile"
  users,'' in \emph{Proc. of IEEE INFOCOM 2004}.

\bibitem{05gass}
R.~Gass, J.~Scott, and C.~Diot, ``Measurements of in-motion 802.11
  networking,'' in \emph{Mobile Computing Systems and Applications, 2006.
  WMCSA'06. Proceedings. 7th IEEE Workshop on}, 2005.

\bibitem{had-mobisys-07}
\BIBentryALTinterwordspacing
D.~Hadaller, S.~Keshav, T.~Brecht, and S.~Agarwal, ``Vehicular opportunistic
  communication under the microscope,'' in \emph{Proc. ofACM MobiSyS 2007}.
  [Online]. Available: \url{http://doi.acm.org/10.1145/1247660.1247685}
\BIBentrySTDinterwordspacing

\bibitem{bal-ccr-08}
\BIBentryALTinterwordspacing
A.~Balasubramanian, R.~Mahajan, A.~Venkataramani, B.~N. Levine, and
  J.~Zahorjan, ``Interactive wifi connectivity for moving vehicles,''
  \emph{SIGCOMM Comput. Commun. Rev.} [Online]. Available:
  \url{http://doi.acm.org/10.1145/1402946.1403006}
\BIBentrySTDinterwordspacing

\bibitem{hare2013dept}
J.~Hare, L.~Hartung, and S.~Banerjee, ``Transparent flow migration through
  splicing for multi-homed vehicular internet gateways,'' in \emph{Proc. of
  IEEE VNC 2013}.

\bibitem{hare2012}
------, ``Beyond deployments and testbeds: Experiences with public usage on
  vehicular wifi hotspots,'' in \emph{Proc. of ACM MobiSyS 2012}.

\bibitem{06-dsrc}
D.~Jiang, V.~Taliwal, A.~Meier, W.~Holfelder, and R.~Herrtwich, ``Design of 5.9
  ghz dsrc-based vehicular safety communication,'' \emph{Wireless
  Communications, IEEE}, 2006.

\bibitem{bal-mobisys-10}
\BIBentryALTinterwordspacing
A.~Balasubramanian, R.~Mahajan, and A.~Venkataramani, ``Augmenting mobile 3g
  using wifi,'' in \emph{Proc. of ACM MobiSys 2010}, 2010. [Online]. Available:
  \url{http://doi.acm.org/10.1145/1814433.1814456}
\BIBentrySTDinterwordspacing

\bibitem{iperf}
\BIBentryALTinterwordspacing
{NLANR/DAST} : Iperf - the {TCP/UDP} bandwidth measurement tool. [Online].
  Available: \url{http://iperf.fr/}
\BIBentrySTDinterwordspacing

\bibitem{rfc2581}
M.~Allman, V.~Paxson, and W.~Stevens, ``Rfc 2581: Tcp congestion control,''
  1999.

\bibitem{rfc2582}
S.~Floyd and T.~Henderson, ``Rfc 2582,'' \emph{The NewReno Modification to
  TCP’s Fast Recovery Algorithm}, 1999.

\bibitem{get-bufferbloat}
\BIBentryALTinterwordspacing
J.~Gettys and K.~Nichols, ``Bufferbloat: Dark buffers in the internet,''
  \emph{Queue}. [Online]. Available:
  \url{http://doi.acm.org/10.1145/2063166.2071893}
\BIBentrySTDinterwordspacing

\bibitem{raychaudhuri2012mobilityfirst}
D.~Raychaudhuri, K.~Nagaraja, and A.~Venkataramani, ``Mobilityfirst: a robust
  and trustworthy mobility-centric architecture for the future internet,''
  \emph{ACM SIGMOBILE Mobile Computing and Communications Review}, 2012.

\end{thebibliography}

\end{document}